# DFCV: A Novel Approach for Message Dissemination in Connected Vehicles using Dynamic Fog

Anirudh Paranjothi[1], Mohammad S. Khan[2], and Mohammed Atiquzzaman[1]

[1]School of Computer Science, University of Oklahoma – Norman, Oklahoma, USA
{anirudh.paranjothi,atiq}@ou.edu
[2] Department of Computing, East Tennessee State University, Johnson City, Tennessee, USA
khanms@etsu.edu

**Abstract.** Vehicular Ad-hoc Network (VANET) has emerged as a promising solution for enhancing road safety. Routing of messages in VANET is challenging due to packet delays arising from high mobility of vehicles, frequently changing topology, and high density of vehicles, leading to frequent route breakages and packet losses. Previous researchers have used either mobility in vehicular fog computing or cloud computing to solve the routing issue, but they suffer from large packet delays and frequent packet losses. We propose Dynamic Fog for Connected Vehicles (DFCV), a fog computing based scheme which dynamically creates, increments and destroys fog nodes depending on the communication needs. The novelty of DFCV lies in providing lower delays and guaranteed message delivery at high vehicular densities. Simulations were conducted using hybrid simulation consisting of ns-2, SUMO, and Cloudsim. Results show that DFCV ensures efficient resource utilization, lower packet delays and losses at high vehicle densities.

## 1 Introduction

Intelligent Transport Systems (ITS) recently developed applications to enhance vehicle safety based on vehicle-to-vehicle (V2V) and vehicle-to-infrastructure communication (V2I). V2V and V2I communication in Vehicular Ad-hoc Network (VANET) depend on Dedicated Short Range Communication (DSRC) [1] which consists of a set of protocols for transmitting messages between vehicles and between vehicles and the roadside infrastructure in a connected vehicular environment. As a result, VANET emerged as the most promising wireless network for a variety of applications from road safety to entertainment.

In a connected vehicular environment, information transmitted among the vehicles in terms of messages. However, many challenges still exist due to the difficulties in deployment and management of resources [2]. In specific, the current techniques for V2V and V2I communications do not provide guaranteed message delivery resulting in messages being dropped before reaching the destination. It is due to an instability of DSRC, arising from the frequency band used by DSRC, as the number of vehicles increases. Furthermore, the current techniques for message dissemination have limitations such as the efficient utilization of resources, delay constraints due to high mobility and unreliable connectivity, and Quality of Service (QoS) [3].

Previous authors used either mobility in vehicular fog computing or cloud computing techniques to solve the instability, resource utilization and QoS problems mentioned above. Wang et al. [4] and Grewe et al. [5] illustrated the possibility of mobility based fog computing for broadcasting information in a vehicular environment. However, the authors did not address various scenarios, including fog-split and fog-merge, in a connected vehicular environment. Moreover, the proposed approach for broadcasting messages [4, 5] has limitations, such as high delay and frequent loss of connectivity. Agarwal et al. [6] discussed techniques to transmit the information between the clouds but creates instability in the cloud if the load increases. In addition, the approach is not suitable for the highly dynamic vehicular environment.

To address the shortcomings, we introduce a fog-based layered architecture, called Dynamic Fog for Connected Vehicles (DFCV) for the dissemination of messages. It consists of two emerging paradigms: 1) fog computing (also known as edge computing), and 2) cloud computing. In contrast to previous methods, DFCV incorporates all possible scenarios for disseminating the messages, including split and merge, in a connected vehicular environment. The *difference* between DFCV and the previous approach [4-6] is the techniques used in deployment and management of resources including broadcasting messages. Previous methods used either cloud computing or mobility in vehicular fog computing to solve the problems, whereas, DFCV uses a three-layered architecture consisting of fog computing and cloud computing techniques, thereby ensuring efficient resource utilization, rapid transmission of messages, decreases in delay and better QoS.

Our *objective* is to lower the delay and to provide guaranteed message delivery at high vehicle densities in a connected vehicular environment. To the best of our knowledge, we are the first to implement the dynamic fog for the dissemination of messages in a connected vehicle environment. The messages are broadcasted to intended recipients using single-hop or multi-hop. Once the message is successfully transmitted, DFCV dismantles the fog. We considered three previously used schemes for comparing with DFCV: 1) Named Data Networking (NDN) with mobility [4], 2) Fog-NDN with mobility [4], and 3) PEer-to-Peer protocol for Allocated REsource (PrEPARE) protocols [7].

The simulations are performed using network simulator (ns-2), Simulation of Urban Mobility (SUMO) and Cloud Simulators (cloudsim). Our *results* lead to an exciting conclusion that the DFCV provides guaranteed message delivery and reduce latency and performs up to 35% better than the current techniques at high vehicle densities and simulation times. The *contributions* of the paper are: 1) developing a framework to broadcast the messages in a connected vehicular environment with guaranteed message delivery, less delay and improved QoS and 2) comparative analysis of the schemes based on the mobility in fog computing.

The rest of this paper is organized as follows: We first discuss related work in Section 2. The proposed system model and the various scenarios involved in it are illustrated in Section 3. Based on the proposed approach, we analyze the performance of our algorithm in Section 4. Performance evaluation of our system is discussed in Section 5. We validate our analysis through extensive simulation in Section 6, before concluding the paper in Section 7.

## 2 Related Work

This section is divided into two main subsections: 1) mobility in vehicular fog computing, and 2) cloud computing.

### 2.1 Mobility in Vehicular Fog Computing

In vehicular fog computing, nodes are equipped with enhanced storage space, communication and computational capabilities at the edge of the internet, usually Road Side Unit (RSU) or base station [8]. Roman et al. [9] illustrated the possibility of fog computing based mobility support, in other words, Mobile Edge Computing (MEC) regarding bringing the resources close to the vehicles in a connected vehicle environment. Furthermore, the authors mentioned about advantages of mobility support in fog including location awareness, availability, low latency, etc. Sun et al. [10] proposed a hierarchical fog computing architecture associated with Virtual Machines (VM) to handle the data generated from terminals such as vehicles, smartphones, etc. at the mobile-edge. Chaudhary et al. [11] explained resource allocation management in fog computing for different application requirements along with the mobility nature of fog users. However, the existing approaches [8-11] has the following limitations: 1) high overhead due to frequent loss of connections and packets and 2) high delay.

### 2.2 Cloud Computing

Botta et al. [12] discussed the integration of cloud computing and the internet of things. Also, they illustrated the service available in the cloud regarding both proprietary and open platforms along with future directions. Paranjothi et al. [13] discussed the performance and outcomes of the mobile cloud in allocation and management resources along with the integration of mobile cloud computing and internet of things. Agarwal et al. [6] proposed a system to perform automatic data placement across geographically distributed data centers. Moreover, the authors discussed various services given by the cloud to its users. The shortcomings of existing approaches [6,12,13] are: 1) high maintenance cost, and 2) delay associated with accessing and allocating resources in the cloud.

## 3 Proposed Architecture of DFCV

In this section, we discuss the system model and the scenarios involved in evaluating DFCV.

### 3.1 System Model

Our proposed fog-based layered architecture, called DFCV, is shown in Fig. 1. DFCV consists of three layers: 1) Terminal layer, 2) Fog layer, and 3) Cloud layer.

**Terminal Layer.** This layer closest to the physical environment and end user. It consists of various devices like smartphones, vehicles, sensors, etc. As the motive of our approach is to broadcast the messages in a connected vehicular environment, only vehicles are represented in the terminal layer. Moreover, they are responsible for sensing the surrounding environment and transmitting the data to the fog layer for processing and storage.

**Fog Layer.** Fog layer is located at the edge of a network. It consists of fog nodes, which includes access points, gateways, RSUs, base station, etc. In DFCV, RSUs and base stations play a major role in disseminating the messages. Fog layer can be static at a fixed location or mobile on moving carriers such as in the vehicular environment. Also, they are responsible for processing the information received from the terminal device and temporarily store it or broadcast over the network.

**Cloud Layer.** The main function of the cloud computing in DFCV is to keep track of the resources allocated to each fog node and to manage interaction and interconnection among workloads on a fog layer, popularly known as fog orchestration.

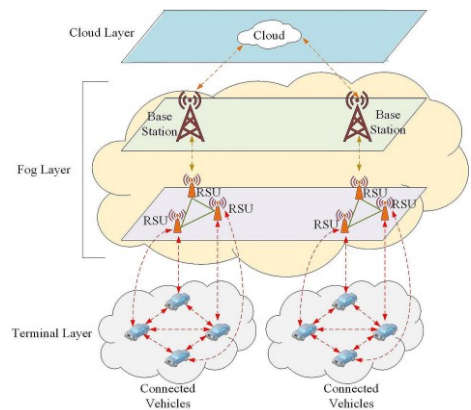

**Fig. 1.** Three layers of our proposed DFCV architecture for dissemination of messages.

**DFCV.** In our approach, the vehicle senses surrounding environment and when it encounters a situation like potholes, road accident, icy road, brake failure, etc. it sends the information to the fog layer or acquires the properties of resources such as RSU and base station from fog layer to broadcast it. As we broadcast the messages in a highly dynamic connected vehicular environment, we coined our approach as DFCV. Our approach supports one to one, one to many and many to many communications.

### 3.2 Scenarios involved in Evaluating DFCV

**Case 1: Split.** Fog split will occur in two scenarios: 1) either the capacity of the DFCV is greater than the threshold capacity (Section 5.2), or 2) the distance between the vehicle increases from the view of the sender, also known as first observer. One possible situation explained in Fig. 2. At time interval ($t1$), the vehicles V3 and V4 connected to RSU 1 and vehicles V1 and V2 connected to RSU 2 respectively. Assume the vehicle V3 need to transmit messages to vehicles V2 and V1. First, messages from V3 transmitted to RSU 1, and then, it relayed to the base station. From the base station, the messages disseminated to the vehicles in it. The cloud monitors the resources used in this activity and keeps track of resources allocated to fog layer.

Consider the same scenario at a time interval ($t2$), since the vehicles are continuously moving in a forward direction and due to frequent topology changes, vehicles V3 and V4 connected to RSU 2 and vehicles V2 and V1 connected to RSU 3 respectively. Here the RSUs are connected to different base stations, and thus, the fog layer splits vertically into two different layers fog layer 1 and fog layer 2. The messages from V3 transmitted to the target base station where the intended recipients (V2, V1) are located using a handshaking technique.

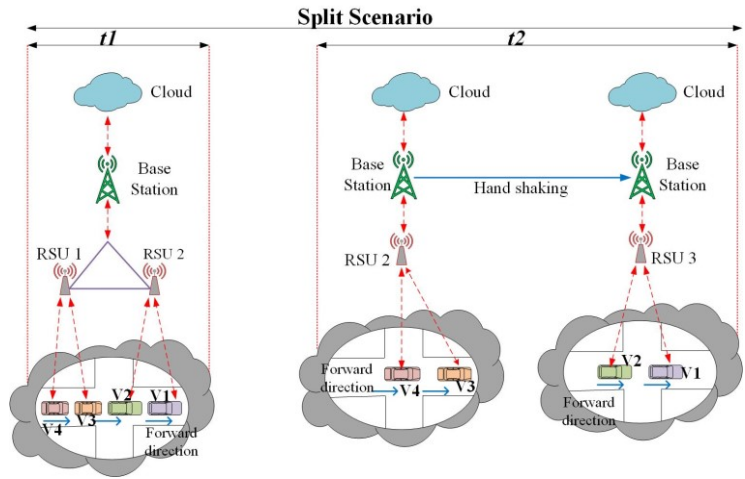

**Fig. 2.** The split scenario of our proposed architecture DFCV for splitting the fog into multiple fogs.

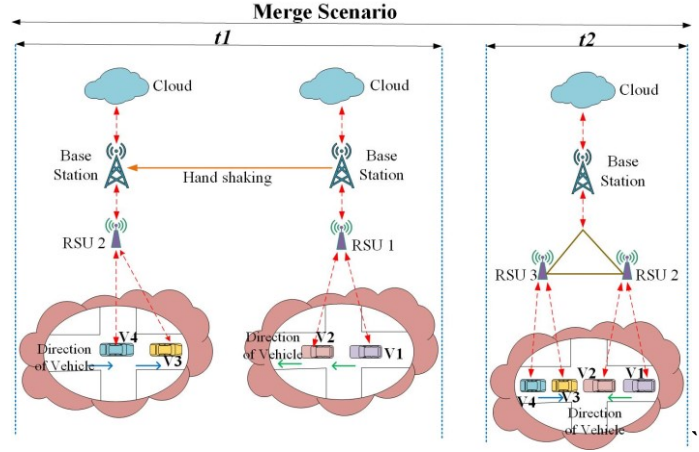

**Fig. 3.** The Merge scenario of our proposed architecture DFCV for merging multiple fogs into a single fog.

**Case 2: Merge.** Fog merge will occur in two scenarios: 1) either the capacity of the DFCV is lesser than the threshold capacity, or 2) the distance between the vehicles is lesser than the minimum distance. At time interval ($t1$), consider the vehicles move in both directions. The vehicles V3 and V4 connected to RSU 2 and vehicles V1 and V2 connected to RSU 1 respectively. Since vehicles V3, V4, and V2, V1 located in a region of different base stations, it gets connected to two different fog layers; fog layer 1 and fog layer 2 as represented in Fig. 3. Assume the vehicle V1 need to transmit messages to the vehicles V3 and V4 respectively. First, messages from V1 transmitted to RSU 1and then relayed to the base station. Since the intended recipients associated with the different base stations, a handshake needs to be performed between them to receive the transmitted message.

Assume the same scenario at a time interval ($t2$). Since the vehicles are continuously moving in both directions, and due to frequent topology changes, vehicles V3 and V4 belong to fog layer 2 and vehicles V2 and V1 belong to fog layer 1 are merged as they come close to each other and forms a single fog layer (fog layer1). The DFCV destroys the fog after successful transmission of messages.

## 4    Proposed Algorithm of DFCV

This section describes our proposed algorithm (DFCV). It is implemented in each cell with the help of fog and contains the information of all the vehicles in the given scenario including the location. The notations used in DFCV are represented in Table 1.

DFCV aims to transmit the messages to the neighboring vehicles using fog computing technique. It mainly concentrates on merge and split scenarios as discussed in Section 3.2. The split is a primitive operation performed by DFCV using *split()* function. The steps are as follows:  First, the distance between the vehicles is calculated using the *distance()* function. It is calculated based on the distance from the sender, and then, the capacity of the DFCV is determined using *th_cap()* function based on the equation formulated in Section 5.2. The split accomplished when the distance exceeds the minimum distance ($d_{min}$) or the capacity of the DFCV ($f_c$) surpass the threshold capacity. Here, a single fog will split into two parts. After the split, messages are relayed to the base station with the help of the RSU and send() function is used to send the input message to the vehicles in a corresponding base station ($bs_i$).

```
Algorithm: DFCV (input_msg, veh_rec)

for all (veh_send ∈ bs_i)
        for all (v ∈ c)
                calculate distance()
                calculate th_cap()
        if (distance > d_min || f_c > th_cap)
                split (v ∈ c)
                split fog_layer
                fog_layer ← v
                bs_i ← send (input_msg)
                veh_rec ← bs_i
        else
                merge (v ∈ c)
                merge fog_layer
                fog_layer ← v
                bs_i ← send (input_msg)
                veh_rec ← bs_i
print message transmitted to an intended recipient(s)
        end if
end for
end for
```

**Table 1.** Notations used in DFCV algorithm.

| Variables | Purpose |
|---|---|
| $veh_{send}$ | Set of the vehicle(s) that need to transmit messages |
| $veh_{rec}$ | Intended recipient(s) |
| $input\_msg$ | Input message from $veh_{send}$ to $Veh_{rec}$ |
| $d_{min}$ | The minimum distance between the vehicles |
| $f_c$ | DFCV capacity |
| $bs_i$ | Base station associated with $veh_{send}$ |
| $v \in c$ | Set of vehicles belongs to the communication range of base station |

Merge is another primitive operation performed by DFCV using *merge()* function based on the following constraints: the distance is lesser than the minimum distance ($d_{min}$), or the capacity of the DFCV ($f_c$) is lesser than the threshold capacity (Section 3.2). It combines two or more fog layers under the same base station ($bs_i$) into a single fog layer. Then, the messages are broadcasted to the neighboring vehicle using *send()* function.

## 5  Performance Evaluation

The performance of DFCV is evaluated using two different analyses: 1) analysis of DFCV and 2) capacity of DFCV. Each analysis is formulated and explained briefly in the following subsections.

### 5.1  Analysis of DFCV

In this analysis, we calculated the probability of failure. Failure of the system can occur due to loss of connectivity or a resource, insufficient capacity of fog, and excessive delays, etc. The probability of system failure ($P_{sysfail}$) is calculated by:

$$P_{sysfail} = \sum_{i=0}^{n_{v,tmax}} \binom{n_{v,tmax}}{i} d_f^i (1-d_f)^{n_{v,tmax}-i} \qquad (1)$$

Where $n_v$, is the number of vehicles in the fog, $tmax$ is the maximum time taken by the vehicles to get connected, and $d_f$ is the probability of success in the fog. Like Quality of Service (QoS), the probability of system failure contributes to the performance of the system. A minimum number of failures leads to the maximum performance of the fog.

### 5.2  Capacity of DFCV

In this analysis, the capacity of DFCV ($f_c$) calculated, and compared with the threshold capacity of the fog ($th_{cap}$). The threshold capacity is calculated based on the resources allocated to the fog node. If the DFCV capacity ($f_c$) is less than the threshold capacity ($th_{cap}$), the communication link established between the vehicles located in that region. Otherwise, a new fog created. The capacity of DFCV is given by:

$$f_c = \frac{n_v}{t_v} \qquad (2)$$

$$\begin{cases} f_c \leq th_{cap}, \text{establish a communication link between vehicles} \\ f_c > th_{cap}, \text{split the fog} \end{cases}$$

Where $f_c$ is the DFCV capacity, $n_v$ is the number of vehicles connected to the fog; $t_v$ is the total vehicles located in the region, $th_{cap}$ is the threshold capacity of the fog.

# 6    Simulation Results

The simulation results to evaluate DFCV are divided into two subsections: 1) simulation setup, explains the various parameters considered for the simulation of DFCV, 2) simulation results, depicts the outcome of our experiments performed using various simulators.

## 6.1    Simulation Setup

Simulation of DFCV algorithm (Section 4) is accomplished based on the architecture and scenarios discussed in Section 3. The analyses of DFCV illustrated in Section 5 are used in simulation to measure the stability of the system (Section 5.1) and to specify when the fog split should occur (Section 5.2) in a given scenario.To simulate the trace of vehicles movements, we used open source traffic simulator SUMO. The output of the SUMO simulator (i.e., the trace of vehicles) given as input to the ns-2 simulator. NS-2 is a discrete event simulator, consisting of many modules to perform the simulation.  Following modules are considered for simulation: 1) node deployment model for dynamic placement of nodes, 2) node mobility model for dynamic network topologies, 3) wireless signal propagation model for transmitting radio waves between the vehicles, and 4) packet loss model to identify the number of packets dropped in transmission. Finally, the cloudsim simulator is used to deploy a cloud to monitor the resources allocated to the fog. The simulations are performed based on the parameters, represented in Table 2.

**Table 2.** Parameters used in simulation of DFCV.

| Parameters | Value |
|---|---|
| Road length [m] | 1000 |
| Number of vehicles [#] | 40-240 |
| Number of lanes [#] | 4 |
| Vehicle speed [mph] | 30-65 |
| Transmission range [m] | 300 |
| Message size [bytes] | 256 |
| Simulator used | ns-2, SUMO, cloudsim |
| Data rate [bps] | 2M |
| Technique used | Multi-hop, Fog, Cloud |
| Protocol | IEEE802.11p |

## 6.2    Simulation Results

We performed simulation in two parts, 1) urban scenario and 2) highway scenario. The urban scenario shows the results of the simulation in an urban environment where splitting or merging takes place less frequently, and the highway scenario shows the results of a simulation performed during the fog split in a highway environment (Fig. 2). As mentioned in Section 3.2, fog split takes place when the distance between the vehicle increases or the capacity of DFCV increases in a region.

**Urban Scenario.** Simulation of DFCV in an urban scenario is performed based on the metrics such as 1) end-to-end delay, 2) probability of message delivery, and 3) collision ratio. During the simulation, DFCV is analyzed based on the equation formulated in Section 5.1 to ensure the stability of the system.

*End-to-end Delay.* The DFCV is aware of the location of the intended recipients; hence, it reduces the time taken for an initial setup across a network from source to destination and disseminate the messages much quicker than existing approaches such as fog-NDN with mobility, NDN with mobility and PrEPARE protocols. Thus, the end-to-end delay of the DFCV is relatively low, represented in Fig. 4a. The end-to-end delay is calculated against the number of vehicles, and it increases as the number of users increases in the system due to a large number of messages need to be delivered within a specific time interval.

*Probability of Message Delivery.* In our approach, we considered the probability of message delivery with respect to the number of vehicles. For each user, the probability of message delivery is distributed in the range of (0-1), as represented in Fig. 4b. From the figure, we can observe that the probability of message delivery decreases marginally as the number of users increases due to the increase in load on the fog. Moreover, DFCV outperforms existing approaches at high vehicle densities.

*Collision Ratio.* To observe the number of packets colliding before reaching the destination, we performed this experiment at a time interval (t) and observed that the collision ratio of our approach is lower at high vehicle densities, and it increases slightly as-the number of users increases in the system, as shown in Fig. 5(a). It is due to the additional packets generated being more likely to encounter another packet and resulting in a collision.

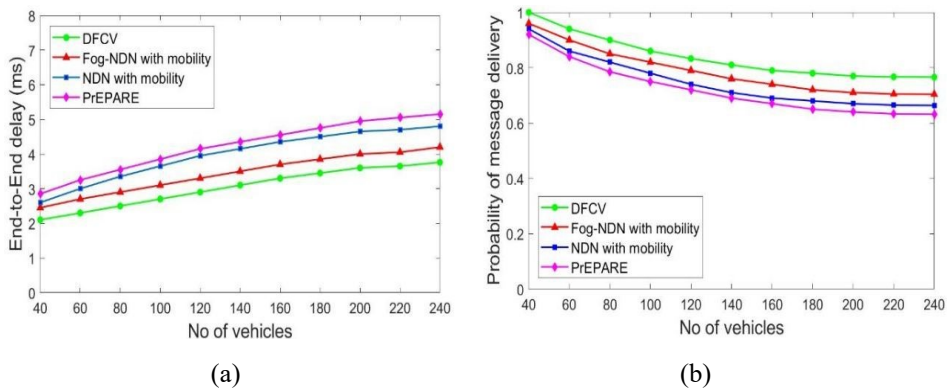

(a)  (b)

**Fig. 4**. Simulation results of DFCV: a) comparison of end-to-end delay in an urban scenario, b) comparison of the probability of message delivery in an urban scenario.

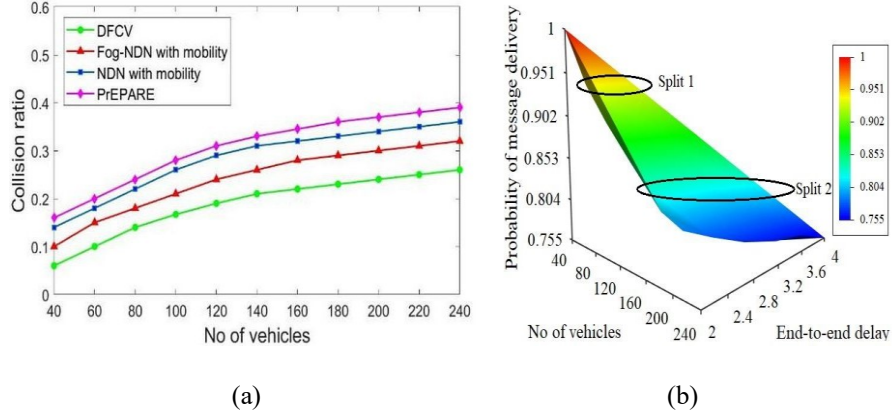

**Fig. 5.** Simulation results of DFCV: a) comparison of collision ratio in an urban scenario, b) split condition in a highway scenario

**Highway Scenario.** Simulation of DFCV in a highway scenario (split condition) is performed based on the metrics discussed in an urban scenario. As the vehicles are moving at high speed, handover occurs frequently, leading to the fog split situation, as represented in Fig. 2. We displayed two possible areas of the split during the simulation in Fig. 5(b). It is calculated based on the equation formulated in Section 5.2. If the split occurs in the yellow region (split 1), the probability of message delivery and end-to-end delay of DFCV is not affected due to the fewer number of vehicles. Whereas, if the split occurs in the blue region (split 2), as the number of vehicles is higher, the end-to-end delay increases when compared to the split 1. However, the performance of DFCV does not deteriorate due to fog split. Furthermore, DFCV yields better performance when compared to existing approaches such as fog-NDN with mobility, NDN with mobility and PrEPARE protocols.

## 7    Conclusion

We studied Connected Vehicle challenges, such as poor resource utilization, increase in delays and frequent vehicle disconnection notably in dense vehicle regions. To address these problems in a connected vehicle environment, we used two emerging paradigms, fog computing, and cloud computing. In this paper, we proposed a novel approach, called DFCV, which ensures less delay and guaranteed message delivery to nearby vehicles. DFCV also supports one to one, one to many, and many to many communications between vehicles. We have analyzed the probability of message delivery, end-to-end delay and the collision ratio by modeling buffers at vehicles and performed simulation using ns-2, SUMO, and cloudsim simulators. The results showed that DFCV is robust, efficient and provides the best performance at all vehicle densities for a number of current schemes available in the literature.

**Acknowledgements.** This work is supported by a grant (RDC # 18-023sm) awarded to the second author from East Tennessee State University.

## References

1. Outay, F., Kammoun, F., Kaisser, F., Atiquzzaman, M.: Towards Safer Roads through Cooperative Hazard Awareness and Avoidance in Connected Vehicles. 31st International Conference on Advanced Information Networking and Applications Workshops (WAINA), pp. 208-215. IEEE, Taipei, China (2011).
2. Hasan, A., Hossain, M., Atiquzzaman, M.: Security threats in vehicular ad hoc networks. 5th International Conference on Advances in Computing, Communications, and Informatics (ICACCI), pp. 404-411. IEEE, Jaipur, India (2016).
3. Paranjothi, A., Khan, M., Nijim, M., Challoo, R.: MAvanet: Message authentication in VANET using social networks. 7th Annual Ubiquitous Computing, Electronics & Mobile Communication Conference (UEMCON), pp. 1-8. IEEE, New York, USA (2016).
4. Wang, M., Wu, J., Li, G., Li, J., Wang, S.: Toward mobility support for information-centric IoV in the smart city using fog computing. In: 5th International Proceedings of the Smart Energy Grid Engineering Conference (SEGE), pp. 357-361. IEEE, Oshawa, Canada, (2017).
5. Grewe, D., Wagner, M., Arumaithurai, M., Psaras, I., Kutscher, D.: Information-centric mobile edge computing for connected vehicle environments: Challenges and research directions. In: 5th International Proceedings of the Workshop on Mobile Edge Communications, pp 7-12. ACM, California, USA (2017).
6. Agarwal, S., Dunagan, J., Jain, N., Saroiu, S., Bhogan. H.: Volley: Automated data placement for geo-distributed cloud services. In: 7th International Proceedings of the Networked System Design and Implementation Conference, pp. 1-16. USENIX, California, USA (2010).
7. Meneguette, R., Boukerche, A.: Peer-to-Peer Protocol for Allocated Resources in Vehicular Cloud Based on V2V Communication. In: 15th International Proceedings of the Wireless Communications and Networking Conference, pp 1-6. IEEE, California, USA (2017).
8. Ni, J., Zhang, A., Lin, X., Shen, X.: Security, privacy, and fairness in Fog-based vehicular crowdsensing. IEEE Communications Magazine 55(6), 146-152 (2017).
9. Roman, R., Lopez, J., Mambo, M.: Mobile edge computing, fog et al.: A survey and analysis of security threats and challenges. Future Generation Computer Systems 78(2), 680-698 (2018).
10. Sun, X., Ansari, N.: EdgeIoT: Mobile edge computing for the Internet of Things. IEEE Communications Magazine 54(12), 22-29 (2016).
11. Chaudhary, R., Kumar, N., Zeadally, S.: Network Service Chaining in Fog and Cloud Computing for the 5G Environment: Data Management and Security Challenges. IEEE Communications Magazine 55(11), 114-122 (2017).
12. Botta, A., De Donato, W., Persico, V., Pescapé, A.: Integration of cloud computing and internet of things: a survey. Future Generation Computer Systems 56(1), 684-700 (2016).
13. Paranjothi, A., Khan, M., Nijim, M.: Survey on Three Components of Mobile Cloud Computing: Offloading, Distribution and Privacy. Journal of Computer and Communications 5(1), 1-31 (2017).